\documentclass{aa} 
\usepackage{hyperref}
\defcitealias{b2021discovery}{LEMC}
\defcitealias{2022gaiaapsisII}{APSIS-II}

\usepackage{lscape}

\usepackage{graphicx}
\usepackage{soul}
\usepackage{txfonts}
\begin{document}

   \title{Emission line star catalogues post-Gaia DR3}

   \subtitle{A validation of Gaia DR3 data using LAMOST OBA emission catalogue}
   
   \author{B. Shridharan \inst{1}\thanks{e-mail: shridharan.b@res.christuniversity.in}, 
           Blesson Mathew \inst{1},
          Suman Bhattacharyya$^{1}$,
          T. Robin$^{1}$,
          R. Arun$^{2}$,
          Sreeja S Kartha$^{1}$,
          P. Manoj$^{3}$,
          S. Nidhi$^{1}$,
          G. Maheshwar$^{2}$,
          K. T. Paul$^{1}$,
          Mayank Narang$^{3}$,
          T. Himanshu$^{3}$\\}
    \authorrunning{Shridharan et al. }
   \institute{Department of Physics and Electronics, CHRIST (Deemed to be University), Hosur Main Road, Bangalore, India
   \and Indian Institute of Astrophysics, Koramangala, Bangalore, India
   \and Tata Institute of Fundamental Research, Homi Bhabha Road, Mumbai, India}

   \date{Received xxxx; accepted xxxx}

 
  \abstract
   {}
   {Gaia DR3 and further releases have the potential to identify and categorise new emission-line stars in the Galaxy. We perform a comprehensive validation of astrophysical parameters from Gaia DR3 with the spectroscopically estimated emission-line star parameters from LAMOST OBA emission catalogue.}
   {We compare different astrophysical parameters provided by Gaia DR3 with those estimated using LAMOST spectra. By using a larger sample of emission-line stars, we perform a global polynomial and piece-wise linear fit to update the empirical relation to convert Gaia DR3 pseudo-equivalent width to observed equivalent width, after removing the weak emitters from the analysis. }
   {We find that the emission-line source classifications given by DR3 is in reasonable agreement with the classification from LAMOST OBA emission catalogue. The astrophysical parameters estimated by \texttt{esphs} module from Gaia DR3 provides a better estimate when compared to \texttt{gspphot} and \texttt{gspspec}. A second degree polynomial relation is provided along with piece-wise linear fit parameters for the equivalent width conversion. We notice that the LAMOST stars with weak H$\alpha$ emission are not identified to be in emission from BP/RP spectra. This suggests that emission-line sources identified by Gaia DR3 is incomplete. In addition, Gaia DR3 provides valuable information about the binary and variable nature of a sample of emission-line stars.}
  {}


   \maketitle
%

\section{Introduction}

Emission-line stars (ELS) are a class of objects with emission lines, particularly H$\alpha$, at 6563 \AA~in the spectrum. They also exhibit physical processes such as stellar winds, jets or outflows, and/or mass accretion through the circumstellar disc. The hot ELS are classified mainly into main-sequence Classical Ae/Be (CAe/CBe; \citealp{rivinius2013classical}) and pre main-sequence (PMS) Herbig Ae/Be (HAeBe; \citealp{waters1998ARA&A..36..233W}) based on its evolutionary stage. Many large sky surveys such as 2MASS \citep{2masscutri2003}, WISE \citep{wisecutri2012vizier}, IPHAS \citep{iphasdrew2005int}, etc., have improved the ELS research by providing precise photometric measurements which are used to classify the ELS into various categories \citep{koenig2014classification, witham2008iphas}.  

The Gaia Data Release 3 (Gaia DR3) catalogue represents a substantial advance in Galactic stellar astronomy. Gaia DR3 \citep{Vallenari2021A&A...649A...1G} builds on previous releases by improving the quality of previously released data and introducing entirely new data products, such as mean dispersed BP/RP spectra from spectro-photometry and radial velocity spectra (RVS), in addition to their integrated photometry in $G_{BP}$, $G_{RP}$, and the white light G-band published in Gaia EDR3 \citep{2022arXiv220606143D}. Gaia BP/RP and/or RVS spectra is now available for sources with G < 19 mag, and astrophysical parameters for sources with G < 17.6 mag.

The previous Gaia releases played a pivotal role in identifying and studying new populations of ELS in the Galaxy. Some notable examples are the selection of 11,000 high confidence PMS from Sco OB2 association \citep{2019A&A...623A.112D}, and understanding the dynamics of young stellar objects (YSOs) in the Vela OB association \citep{2019A&A...626A..17C}. The Spitzer/IRAC Candidate YSO (SPICY) catalogue was compiled from the YSO candidates identified using the high-quality astrometric data from Gaia EDR3 along the Galactic midplane \citep{2022arXiv220604090K}. More homogeneous studies on the stellar parameters of YSOs were carried out by \cite{2019yCat..51570159A} and \cite{2020MNRAS.493..234W} using Gaia DR2, and \cite{2021A&A...650A.182G} and \cite{2022ApJ...930...39V} using Gaia EDR3. 
Even though Gaia has extensively improved stellar parameters of the previously known ELS in the Milky Way, the unavailability of H$\alpha$ emission measurements for the Gaia sources hindered the classification of more ELS.

The Large sky Area Multi-Object fibre Spectroscopic Telescope (LAMOST) has observed and catalogued 10,431,197 spectra of astronomical sources in their latest DR7 data release. Due to the availability of such a large database of spectra, the number of newly identified ELS has improved. \cite{2016RAA....16..138H} identified 10,436 early-type ELS using LAMOST DR2 and studied various H$\alpha$ profiles. \citet[hereafter, called as \citetalias{b2021discovery}]{b2021discovery} compiled a catalogue of 3339 hot ELS from 451,695 O, B, and A-type spectra from the LAMOST DR5 release. After careful spectral type re-estimation, they reported 1088 CBe, 233 CAe, and 56 HAeBe stars based on the analysis of optical/IR magnitudes and colours. This makes it one of the largest homogeneous ELS catalogue with a thorough classification using spectroscopy and available photometry. More recently, \cite{2022ApJS..259...38Z} identified 25,886 early-type ELS from LAMOST DR7. Even though the number of ELS objects increased with such large spectroscopic surveys, they cannot be classified accurately unless astrometric and photometric data are available. Hence, the field of ELS improves when the large spectroscopic surveys and all-sky astrometric surveys progress in tandem. This is achieved by the recently released Gaia DR3 data which provides astrometric, photometric, and spectroscopic parameters for more than 200 million objects. There is no doubt that the DR4 and further releases will greatly improve the ELS research. 

As the first step in this direction, we compare the new dataset released by Gaia DR3 with a previously existing, well-characterised spectroscopic catalogue. In this work, we aim to provide an external validation for the astrophysical parameters and to improve our ELS catalogue with newly available data from Gaia DR3.

\section{Data analysis and Results}
\label{sec:analysis}
 We use the 3339 ELS from \citetalias{b2021discovery} and queried various DR3 tables using the source identifier from EDR3. The query was made using ADQL facility in the Gaia archive~\footnote[1]{\href{https://gea.esac.esa.int/archive/}{https://gea.esac.esa.int/archive/}}. We explore the different datasets that Gaia provides with its new release. 

\begin{figure}
    \centering
    \includegraphics[width=1\columnwidth]{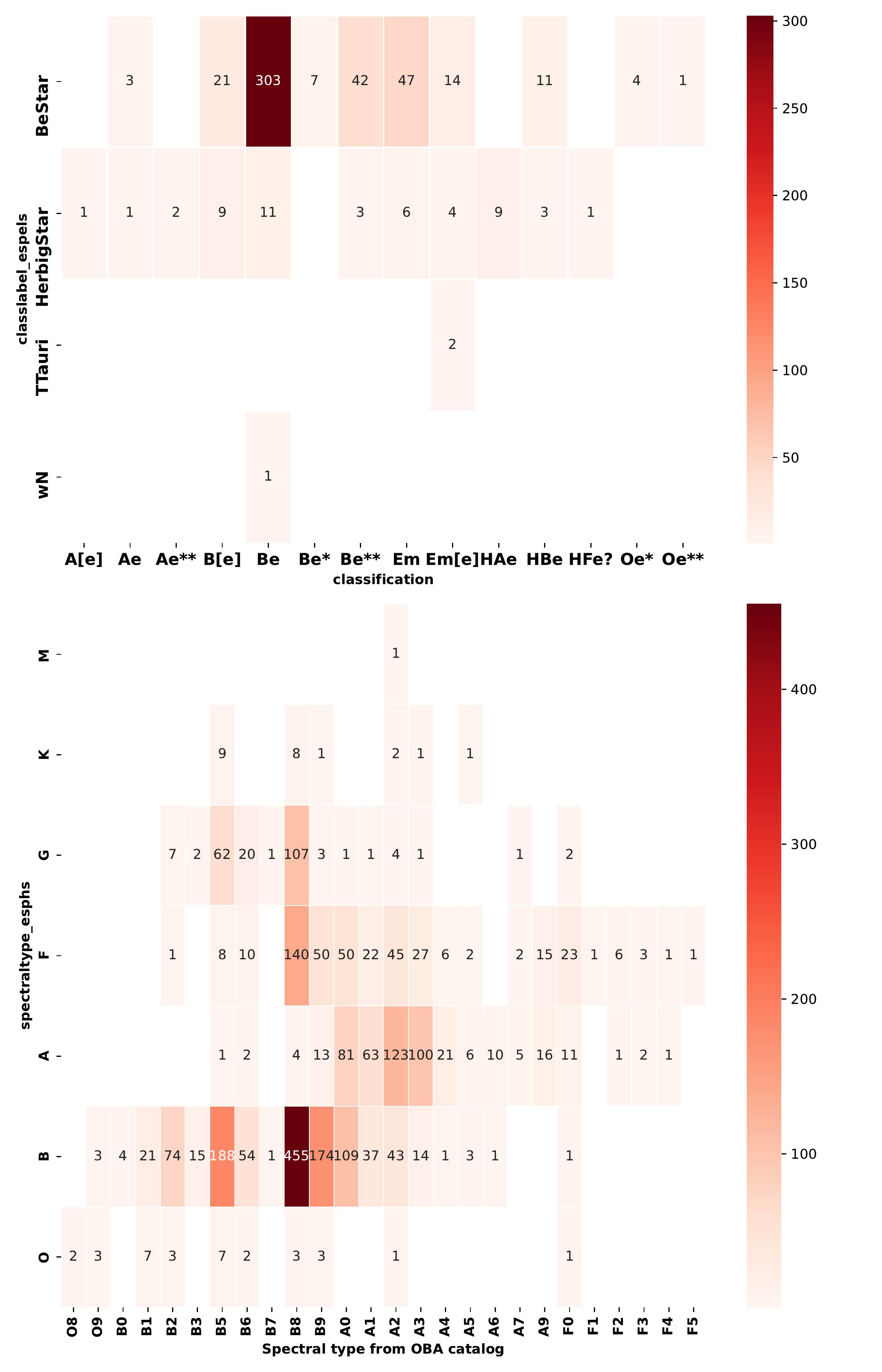}
    \caption{(Top) A heatmap representation of comparison between \citetalias{b2021discovery} classification and \texttt{spectraltype\_esphs} provided by Gaia DR3. (Bottom) A comparison between spectral type from \citetalias{b2021discovery} and \texttt{spectraltype\_esphs} provided by Gaia DR3. The number statistics for each category is provided inside each cell and a color bar is given for the reference.}
    \label{fig:combined_heatmap}
\end{figure}

\subsection{Classification and astrophysical parameters from Gaia DR3}

\begin{figure*}
    \centering
    \includegraphics[height=11cm, width=19cm]{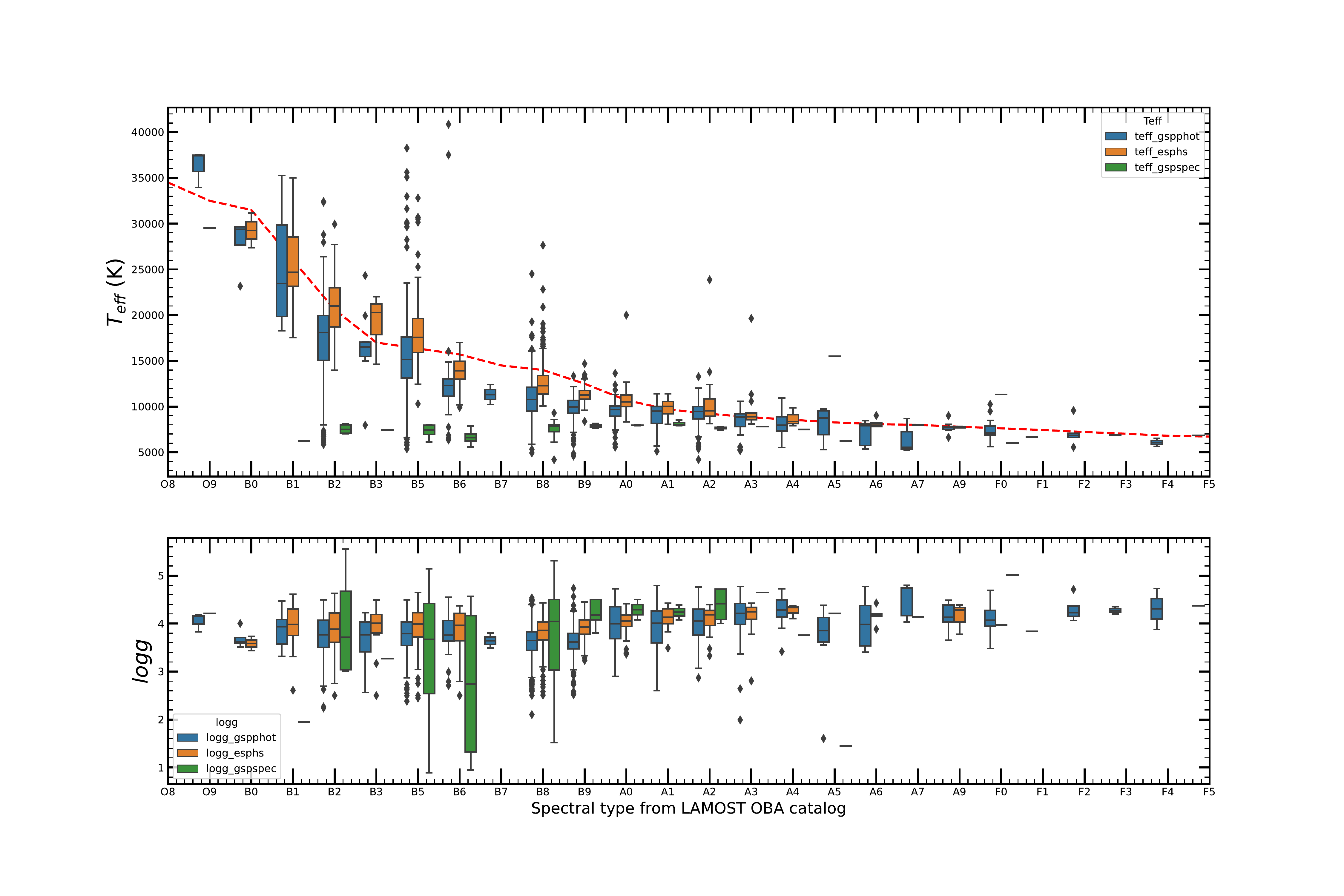}
    \caption{(Top) The distribution between different $T_{eff}$ values provided by Gaia DR3 and the spectral type estimated in \citetalias{b2021discovery} is shown as a boxplot. The red dashed line represent the $T_{eff}$ vs spectral type calibration relation from \cite{2013Pecaut}. (Bottom) The distribution between  $logg$ values of stars from Gaia DR3 and the spectral type estimated in \citetalias{b2021discovery} is shown as a boxplot.}
    \label{fig:Teff_plot_comparision.pdf}
\end{figure*}

The \texttt{gaiadr3.astrophysical\_parameters} table provides plenty of information using the BP/RP spectra, the details of which can be found in \citet[hereafter APSIS-II]{2022gaiaapsisII}. The comparison between the sub-classification of ELS reported in \citetalias{b2021discovery} with the classification done using the `Extended Stellar Parametrizer for Emission-Line stars (ESP-ELS)' module of Gaia DR3 (mentioned as \texttt{classlabel\_espels}), for a sample of 506 stars, is shown as a heatmap in the top panel of Figure \ref{fig:combined_heatmap}. The bottom panel of Figure \ref{fig:combined_heatmap} shows the heatmap of the spectral type comparison between 3109 ELS from LEMC with those estimated from the `Extended Stellar Parametrizer for Hot Stars (ESP-HS)' module in Gaia DR3 (denoted as \texttt{spectraltype\_esphs}).

\begin{figure*}
    \centering
    \includegraphics[height=7cm, width=17cm]{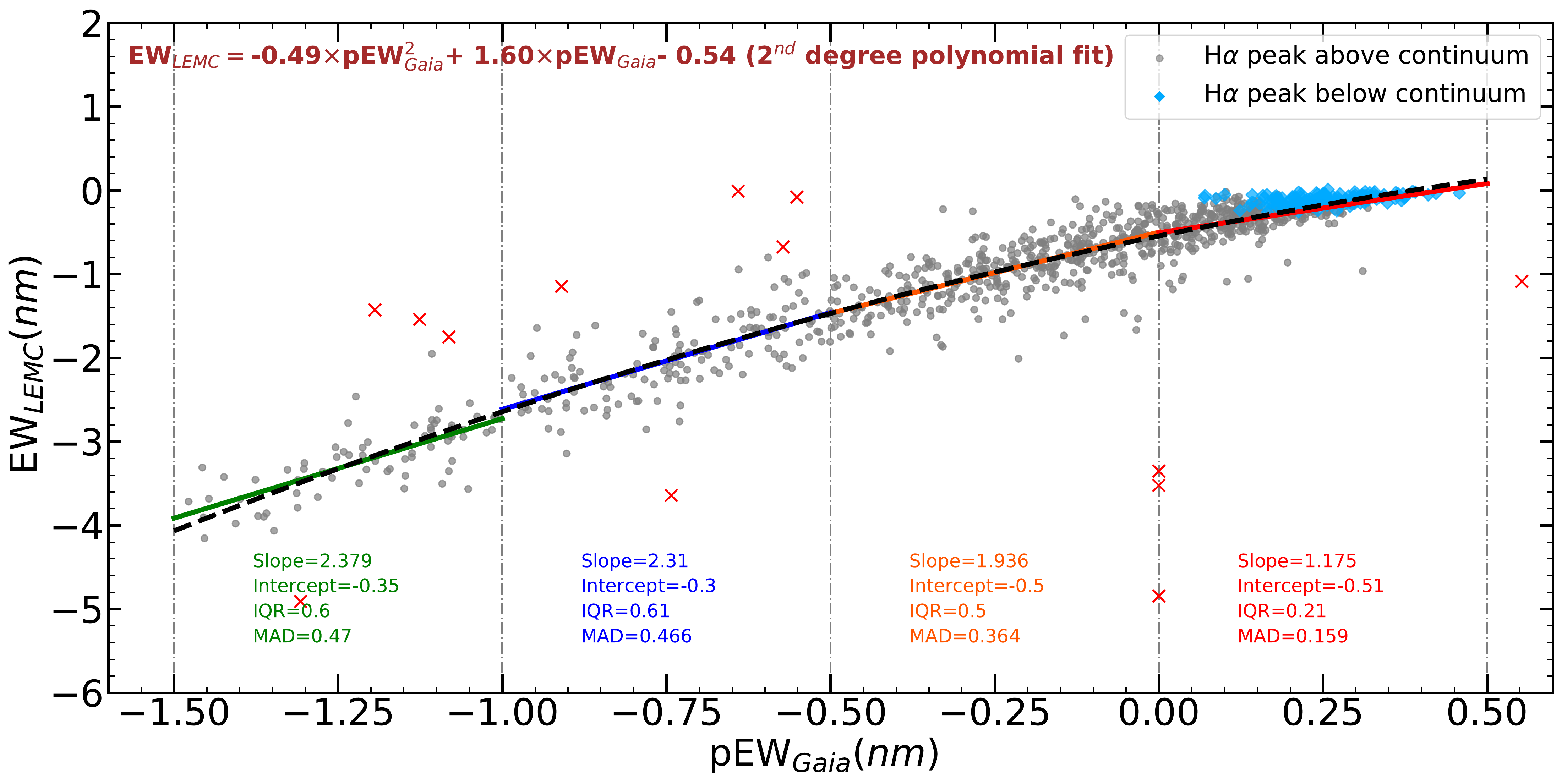}
    \caption{The figure shows a scatter plot between pEW provided by Gaia DR3 and EW estimated for LEMC CBe stars. The red crosses shows the outliers based on a normal distribution analysis. The grey filled circles show the stars with H$\alpha$ peak above the continuum and light blue diamonds with H$\alpha$ peak below the continuum. A global second degree polynomial fit is shown in black dashed lines along with the equation at the top left of the plot}. Further, the piecewise fit within each interval is shown along with the fit parameters in corresponding colours. Negative EW values denote lines in emission.
    \label{fig:EQW estimate}
\end{figure*}

From the figure (Figure \ref{fig:combined_heatmap}; top panel), we see that the classification provided by \citetalias{b2021discovery} and \texttt{classlabel\_espels} DR3 matches well. Of 315 CBe stars with Gaia DR3 estimates, 303 (96 \%) stars are classified as \lq BeStar\rq, 11 (4 \%) stars as \lq HerbigStar\rq and 1 star as \lq wN \rq by Gaia DR3. The quality of \texttt{classlabel\_espels} classification is given by the \texttt{classlabel\_espels\_flag}, where \texttt{classlabel\_espels\_flag} <= 2 denotes a probability larger than 50\%. Interestingly, the 11 stars which are classified as Herbig stars are having quality flag \texttt{classlabel\_espels\_flag} >= 4. For the 303 stars classified as `BeStar', 155 stars have \texttt{classlabel\_espels\_flag} <= 2 and 148 have \texttt{classlabel\_espels\_flag} > 2. The sub-sample of 89 stars with unclear classifications in \citetalias{b2021discovery} (Be**, Be*, Em*) \footnote[2]{Be** = LEMC B-type star but no detection in Gaia EDR3 \\
Be* = LEMC B-type star with Gaia EDR3 detection but not in 2MASS \\
Em* = H$\alpha$ emission object for which spectral type could not be calculated \\} can now be classified as \lq Be Star\rq (83) and \lq Herbig Star\rq (6).

The bottom panel of Figure \ref{fig:combined_heatmap} shows the comparison between the spectral type given in \citetalias{b2021discovery} and  those estimated by Gaia DR3, \texttt{spectraltype\_esphs}. It can be seen that stars with \texttt{spectraltype\_esphs}=\lq B\rq, the spectral type estimated are reasonably matching with \citetalias{b2021discovery} spectral types ranging from O (<1\%), B0-B5 (25\%), B5-B9 (57\%) to A0-A5 (17\%).  However, the problem with \texttt{spectraltype\_esphs} can be seen clearly when we consider the stars with \citetalias{b2021discovery} spectral type B8 (767 stars). Of the 767 stars, 255 (33\%) stars are classified by Gaia DR3 to be \texttt{spectraltype\_esphs}=F/G/K. This is a very significant deviation from the accurate spectral type given in \citetalias{b2021discovery}, which was performed through a semi-automated template matching technique. The deviation of 33\% towards later spectral types should be kept in mind before using the \texttt{spectraltype\_esphs} in future studies. A possible explanation for the observed deviation can be the line-of-sight extinction. The 33\% of the B8 stars misclassified by Gaia DR3 as F/G/K have higher extinction values in both Green's 3D dustmap \citep{green20193d} and Gaia DR3 (\texttt{AG-DR3}), whereas the extinction value for the 59\% of B8 stars classified to be B spectral type, is within 0-1 mag. Thus, higher the observed extinction value, higher the chances of Gaia DR3 spectral type estimation being different from the spectral type in \citetalias{b2021discovery}.

Gaia DR3 provides several astrophysical parameters such as $T_{eff}$, $logg$, $V$sin$i$, mass, radius and luminosity based on the BP/RP spectrum. For hot stars and ELS, they have used special modules to estimate these parameters. We compare all the different $T_{eff}$ estimates with our spectral type to identify the best value for hot ELS. It should be noted that spectral type estimates from \citetalias{b2021discovery}, although performed meticulously, have errors of about ±2 subtypes. Figure \ref{fig:Teff_plot_comparision.pdf} shows the distribution of various $T_{eff}$ and $logg$ estimates of ELS available from Gaia DR3 with spectral type estimated in \citetalias{b2021discovery}. It is very evident from  Figure \ref{fig:Teff_plot_comparision.pdf}~(top) that for B-type stars, $T_{eff}$ is significantly underestimated using RVS spectra (\texttt{teff\_gspspec}). Two different modules were used to estimate $T_{eff}$ using BP/RP spectra i.e., \texttt{teff\_gspphot} and \texttt{teff\_esphs}.  Figure \ref{fig:Teff_plot_comparision.pdf}~(top) reveals that the \texttt{teff\_esphs} value matches better when compared to $T_{eff}$ from \cite{2013Pecaut} calibration table and also, it has significantly lower inter-quartile range (IQR) when compared to \texttt{teff\_gspphot}. We notice a large number of outliers in the \texttt{teff\_gspphot} boxplot for each spectral type, which questions its validity. Hence it is clear from our analysis that \texttt{teff\_esphs} provides an better $T_{eff}$ estimate for B-type stars. In addition, there are other $T_{eff}$ estimates available from modules such as \texttt{teff\_gspphot\_marcs}, \texttt{teff\_gspphot\_ob} and \texttt{teff\_gspphot\_a} in the \texttt{gaiadr3.astrophysical\_parameters\_supp} table. An appropriate model selection can be done based on the object of interest.

Similarly, Figure \ref{fig:Teff_plot_comparision.pdf}~(bottom) shows the distribution of $logg$ values for a subsample of \citetalias{b2021discovery} stars. The $logg$ estimate from RVS spectra (\texttt{logg\_gspspec}) shows a large scatter when compared to the $logg$ estiamtes from BP/RP spectra i.e., \texttt{logg\_gspphot} and \texttt{logg\_esphs}, which are distributed in the range 3-4 dex. Since the \citetalias{b2021discovery} sample contains mainly CBe and HAeBe stars, it is fair to expect $logg$ to be within 3-5. Hence we conclude that, when compared to other modules used in Gaia DR3, the ESP-HS module provides accurate astrophysical parameters and can be used for the analysis of OBA stars. According to \cite{2022Fremat_Gaiadr3}, the $V$sin$i$ estimations from the \texttt{vbroad} module degrades noticeably at $T_{eff}$ > 7500 K and $G_{RVS}$ > 10. Therefore, the $V$sin$i$ would be highly inaccurate for our sample of hot ELS stars. Consequently, we did not include $V$sin$i$ analysis in the present study.

\subsection{Comparison of Gaia DR3 pEW with EW from LAMOST spectra}

Gaia DR3 made available pseudo-equivalent width (pEW) measurements of H$\alpha$ for about 235 million sources, which are given in Gaia DR3 \texttt{astrophysical\_parameters} table as  \texttt{ew\_espels\_halpha} parameter. The classification and the ELS catalogue provided by Gaia DR3 are dependent on this pEW calculation. However, due to the low resolution of BP/RP spectra, using pEW solely may not provide a complete list of ELS which can be identified from Gaia DR3. Hence it is important to calibrate pEW values with actual EW measurements carefully. \citetalias{2022gaiaapsisII} provides an empirical relation between pEW and the EW values available from various ELS catalogues in the literature (Figure 21 and Table 3 of  \citetalias{2022gaiaapsisII}). They estimated the slope of the linear fit to be in the range of 2.26 and 2.83,  which can be used to convert pEW to actual H$\alpha$ EW.

\begin{figure}
    \centering
    \includegraphics[height=6cm,width=1.1\columnwidth]{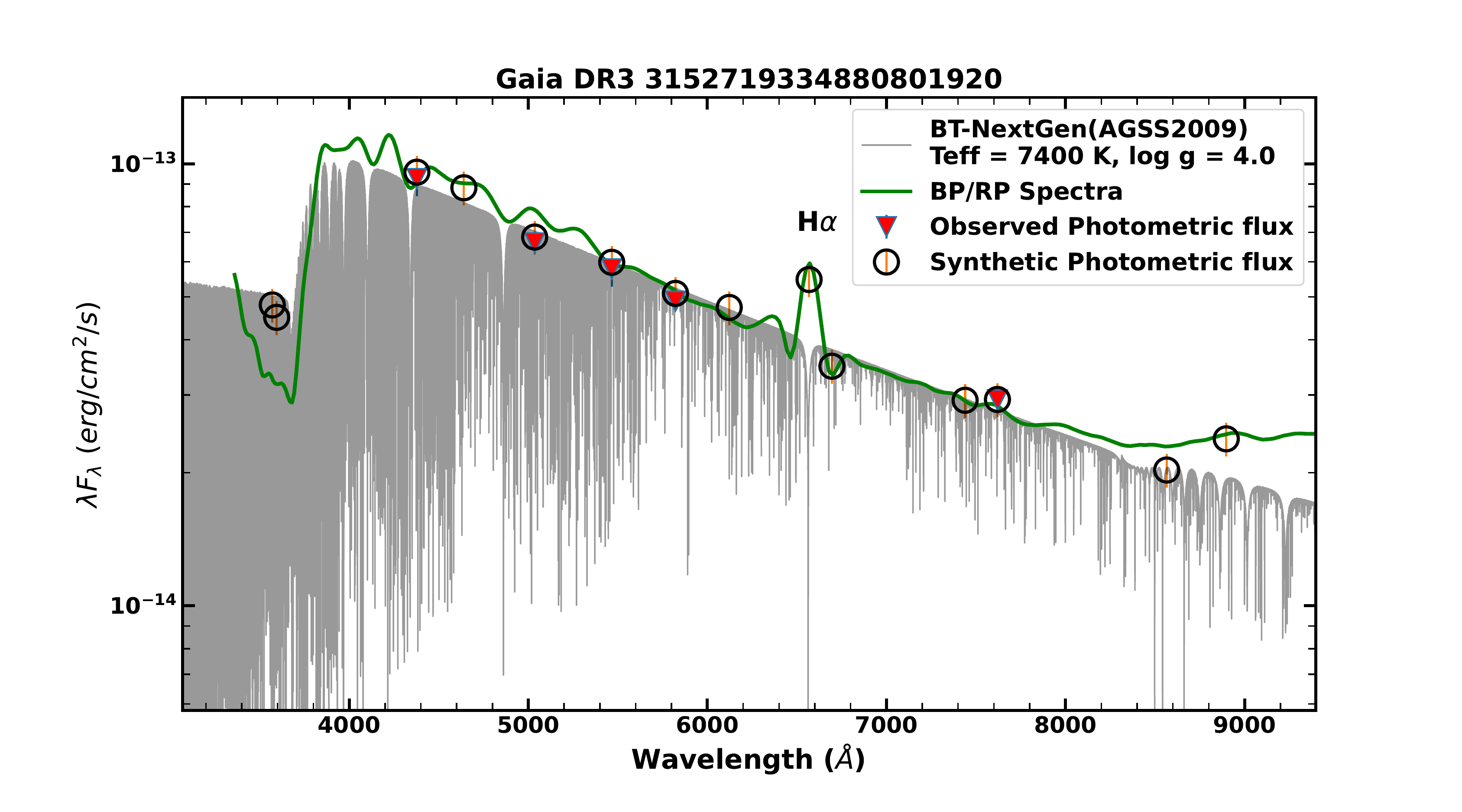}
    \caption{Spectral Energy Distribution of Gaia DR3 3152719334880801920 with $\lambda~F_{\lambda}$ in y-axis and wavelength in x-axis. Red triangles denote the photometric flux available pre-Gaia DR3. Black hollow circles represent the synthetic photometry calculated using Gaia BP/RP continuous spectrum. Gaia DR3 mean sampled BP/RP spectrum are shown in green solid line. The best fit BT-NextGen spectrum are shown in grey.}
    \label{fig:sed_comparison}
\end{figure}

We improve upon this analysis by performing a second degree polynomial fit to a large sample of 1088 CBe stars from \citetalias{b2021discovery}. Even though we have a bigger sample of 3339 ELS, we do not attempt to make a fit with other classes to avoid problems like emission inside the absorption core (CAe stars; \citealp{2021anusha}), the low number statistics (HAeBe) and the contamination from [NII] forbidden lines. We use the sample of 1088 CBe stars from \citetalias{b2021discovery} for which the EW were measured homogeneously using IRAF (Anusha et al., in prep). Stars showing H$\alpha$ emission peak inside the absorption core are shown (light blue diamonds) in Figure \ref{fig:EQW estimate} and were not used in the analysis. We emphasise here that, Gaia DR3 identifies the H$\alpha$ to be in emission only if the emission peak is above the local continuum. Thus, for B-type stars, Gaia DR3 can identify sources as ELS only if the observed EW is greater than 0.5 nm. For A-type stars, the threshold value will only increase, since the H$\alpha$ absorption peaks at A0 spectral type \citep{graycorb2009ssc..book.....G}. Hence the catalogue of ELS provided by Gaia DR3 may not be complete with weak emitters, specifically those with emission peak inside the absorption core. This is a known caveat owing to the very low resolution of BP/RP spectra \citep{gaiaels2008sf2a.conf..499M}. The second degree polynomial relation is shown in Equation \ref{eqn}.  
\begin{equation}
\ EW_{LEMC} ~(nm)~ = -0.54~+~1.60~\times~pEW_{Gaia}~-~0.49~\times pEW{^2}_{Gaia} ~(nm)~
\label{eqn}
\end{equation}
Since we have larger sample of CBe stars when compared to \citet{silaj2010systematic} and \citet{raddi2015deep}, we also perform piece-wise linear fit in intervals of 0.5~nm. The slope and intercept of the linear fit along with IQR ($EW_{LEMC}$) and median absolute deviation (MAD) along $EW_{LEMC}$ axis as a representative of the scatter is given for each interval range. A global  polynomial fit and a piece-wise fit for different intervals of pEW values are shown in Figure \ref{fig:EQW estimate}. As seen from piece-wise linear fit values, the slope gets steeper as we move towards intense emitters. We suggest using the respective slopes and intercepts for calculating observed EWs from each pEW range for hot ELS (Figure \ref{fig:EQW estimate}). However, the users should be aware that the LAMOST and Gaia have obtained the spectra at different epochs; the scatter and deviation of some points can be attributed to intrinsic variability of some CBe stars that range in the orders of days to years \citep{2011MathewBASI, cochetti2021intriguing}. The addition of pEW measurement in DR3 will improve the sample of ELS and can serve as target list for future H$\alpha$ ELS surveys.

\begin{figure}
    \centering
    \includegraphics[width=1\columnwidth]{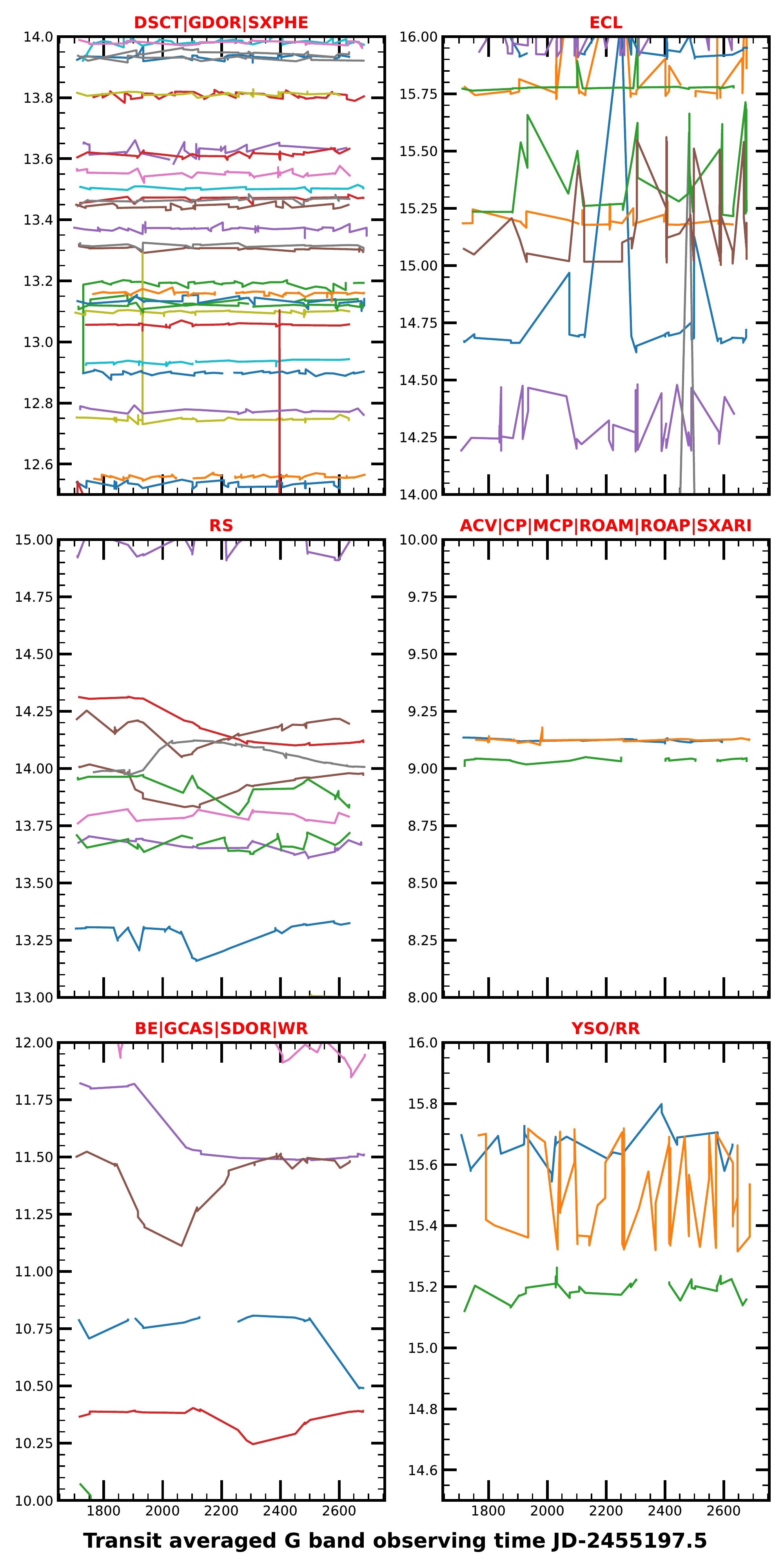}
    \caption{G-band multi-epoch photometry of stars classified as variables. The different subplots show the different variable classes as provided by Gaia DR3 with the class specification shown in red letters. The y-axis is limited to a range of 2 mag to visualize the variability in each class. The description on each variable class is provided in Table 2 of \citet{2022arXiv220606416E}.  }
    \label{fig:gband-variability}
\end{figure}  

\subsection{Synthetic photometry from BP/RP spectra}

For 686 stars in the \citetalias{b2021discovery}, we could not estimate the spectral type due to the low SNR in the bluer region of the LAMOST spectra. Due to the observation strategy of LAMOST DR5, the majority of our sample is towards the galactic anti-center direction (Figure 1 of \citealp{b2021discovery}). Limited photometric survey footprints towards this region restricted us from studying these stars photometrically or using SEDs to estimate their stellar parameters. For our sample of 3339 ELS, 2872 stars have continuous BP/RP spectra. The \texttt{gaiaxpy} package enables the user to calculate the synthetic magnitudes based on the continuous BP/RP spectra from DR3. 
We used the \texttt{gaiaxpy} package to generate Johnson, SDSS, PanSTARRS, and IPHAS  photometric magnitudes for our sample of 2872 stars. %
To show the improvement of creating SED using BP/RP spectra, we present a representative SED which compares the data before and after incorporating Gaia DR3 data, in Figure \ref{fig:sed_comparison}. The SED is fitted with a Python routine used in \cite{2021MNRAS.507..267A} and \cite{2022Bhattacharyya2}. We suggest that for sources with bad quality photometric measurements, synthetic photometry from BP/RP spectra can be used to improve the SED studies.  

\begin{table*}
\small
    \caption{Table containing the non-single stars data from Gaia DR3 for a subset of \citetalias{b2021discovery} stars. SB1 = Spectroscopic Binary}
    \setlength{\tabcolsep}{3pt}
    \begin{tabular}{cccccccc}\hline

    DR3Name & LAMOST ID & Classification & nss\_solution\_type & Period & Inclination & Eccentricity & Center\_of\_mass\_velocity \\  
    & & & & (days) & (degrees) & & (km/s) \\
    \hline
    440218930776209664 & J030719.47+523014.2 & Be*-onlynir & SB1 & 47.53 & - & 0.182 & -57.457 \\
    2685047840736856448 & J212547.34-022251.2 & Ae*-onlynir & SB1 & 195.252 & - & 0.326 & -143.58  \\ 
    4549079418323779712 & J173714.72+160334.7 & F[e]? & SB1 & 8.55 & - & 0.4349 & -15.50\\ 
    760463232938062464 & J112045.14+362535.6 & Ae & SB1 & 6.93 & - & 0.053 & 1.405 \\
    948585824160038912 & J072441.51+404013.1 & Be & SB1 & 8.20 & - & 0.077 & 9.46  \\ 
    666842467830419200 & J074942.34+153117.3 & Em & SB1 & 11.47 & - & 0.25765 & -13.58  \\ 
    277055356579399296 & J043023.15+550408.8 & Be*-onlynir & SB1 & 57.082 & - & 0.01306 & -20.97  \\ 
    2081810716132810368 & J200645.38+435107.9 & Em[e] & Orbital & 573.77 & - & 0.469 & -  \\ \
    3441613167517590400 & J053943.43+265316.2 & Em[e] & EclipsingBinary & 1.48 & 77.59 & 0.0 & - \\ 
    3340108762301888256 & J054241.85+114343.3 & A[e] & EclipsingBinary & 0.876 & 72.08 & 0.0 & - \\ \hline

    \end{tabular}
\label{table: nss}
\end{table*}

\subsection{Non-single stars and variable stars}

One of the major improvements in Gaia DR3 is the classification of 813,687 stars as non-single stars with  orbital binary solutions for 356,132 stars. Massive stars are known to have a binary companion or clustering around it \citep{chini2013massive}. Hence, we use non-single stars catalogue to find the binary stars in our sample. Among our sample of ELS, only 10 have solutions in \texttt{gaiadr3.nss\_two\_body\_orbit} which gives the parameters for spectroscopic and eclipsing binaries. They also provide \texttt{mass\_ratio}, \texttt{eccentricity}, \texttt{inclination}, \texttt{teff\_ratio}, which can be used to characterise binaries. The LAMOST ID along with the parameters from \texttt{gaiadr3.nss\_two\_body\_orbit} is shown in Table \ref{table: nss}. 

Gaia DR3 classified a sample of its sources into different variable categories based on multi-epoch photometry.  From \citetalias{b2021discovery}, 363 stars are classified as variable stars in Gaia DR3. Epoch photometry of variable stars with good quality classification (\texttt{best\_class\_score}  > 0.6) are shown in Figure \ref{fig:gband-variability}. The cause of variability can also be related to the evolving nature of $H\alpha$ emission region and hence, a detailed analysis of these stars will be taken up in a future work. 

\section{Summary}

The newly released Gaia DR3 data will accelerate the field of astronomy as it provides astrophysical parameters for 470,759,263 sources using the mean BP/RP spectra. In that, 2,382,015 sources are classified as hot stars which can increase the number of known CBe and HAeBe stars. As a first step towards achieving this, we compared the astrophysical parameters provided by DR3 with carefully classified OBA-type ELS identified from LAMOST DR5.  

We see that the ELS classification provided by Gaia DR3 as \texttt{classlabel\_elseps} matches reasonably well with our \citetalias{b2021discovery} catalogue. Gaia DR3 also provides new classification and spectral type estimate for stars classified as \lq Em\rq  and \lq Em[e]\rq  in \citetalias{b2021discovery} catalogue. The mismatch between the spectral types provided by Gaia DR3 (\texttt{spectraltype\_esphs}) and \citetalias{b2021discovery} was evident on comparison. The \texttt{spectraltype\_esphs} estimates should be used with caution along with quality flag provided. Gaia DR3 also provides $T_{eff}$ from 3 different modules using both BP/RP spectra and RVS spectra. Based on our comparison of $T_{eff}$ values with spectral types from \citetalias{b2021discovery} catalogue, we see that \texttt{teff\_esphs} values matches well with the theoretical values. The \texttt{teff\_gspspec} values are severely underestimated for early B-type stars.  Similarly,  \texttt{teff\_gspphot} estimate may not be reliable because of the scatter and high number of outliers. We conclude that, \texttt{teff\_esphs} should be used as the $T_{eff}$ estimate for early-type ELS.

We used the sample of 1088 CBe stars from \citetalias{b2021discovery} to perform a global polynomial fit and piece-wise fit analysis to obtain a relation to convert pEW to the actual H$\alpha$ EW. In cases where one needs a more accurate estimate of actual $H\alpha$ EW for a specific range of pEW, the piece-wise slope and intercept values can be used. It should be noted that the weak emitters (with emission peak inside the absorption core) in \citetalias{b2021discovery} have positive pEW values in Gaia DR3. This directly implies the incompleteness of ELS catalogue provided by Gaia DR3. 

We also checked for non-single stars and variable stars present in \citetalias{b2021discovery} catalogue. Among our sample, 10 non-single stars with 7 of them classified as spectroscopic binaries for which various parameters are provided. From \citetalias{b2021discovery}, 363 stars are classified as variables. These H$\alpha$ emitting binaries and variable ELS will be studied in a future work. 

To summarise, this work provides an account of how the data provided by the recent Gaia DR3 can improve the study of ELS. Along with photometry and astrometric measurements, the availability of BP/RP spectra for a large number of sources will increase the number of already known ELS. The astrophysical parameters estimated from the BP/RP and RVS spectra will help to study a large number of ELS with ease.

\begin{acknowledgements}
We would like to thank the Science \& Engineering Research Board (SERB), a statutory body of the Department of Science \& Technology (DST), Government of India, for funding our research under grant number CRG/2019/005380. We thank the Center for Research, CHRIST (Deemed to be University), Bangalore, India, for funding our research under the grant number MRP DSC-1932. This work has made use of data from the European Space Agency (ESA) mission {\it Gaia} (\url{https://www.cosmos.esa.int/gaia}), processed by the {\it Gaia} Data Processing and Analysis Consortium (DPAC, \url{https://www.cosmos.esa.int/web/gaia/dpac/consortium}). Funding for the DPAC has been provided by national institutions, in particular the institutions
participating in the {\it Gaia} Multilateral Agreement. Guoshoujing Telescope (the Large Sky Area Multi-Object Fiber Spectroscopic Telescope LAMOST) is a National Major Scientific Project built by the Chinese Academy of Sciences. Funding for the project has been provided by the National Development and Reform Commission. LAMOST is operated and managed by the National Astronomical Observatories, Chinese Academy of Sciences. We thank the SIMBAD database and the online VizieR library service for helping us with the literature survey and obtaining relevant data.
\end{acknowledgements}

\bibliographystyle{aa}
\bibliography{biblograph}

\end{document}